\documentclass[aps,pra,showpacs,twocolumn,superscriptaddress]{revtex4}
\usepackage{array}
\usepackage{booktabs}
\usepackage{tabu}
\usepackage{dcolumn}
\usepackage{amsmath}
\usepackage{amsfonts}
\usepackage{float}
\usepackage{amssymb}
\usepackage{graphicx,color}
\usepackage[colorlinks={true}]{hyperref}
\usepackage{graphicx}
\usepackage{subfigure}
\usepackage{graphicx}
\usepackage{dcolumn}
\usepackage{bm}

\def\be{\begin{equation}}
  \def\ee{\end{equation}}
\def\bea{\begin{eqnarray}}
\def\eea{\end{eqnarray}}
\def\f{\frac}
\def\n{\nonumber}
\def\l{\label}
\def\p{\phi}
\def\o{\over}
\def\R{\rho}
\def\pa{\partial}
\def\om{\omega}
\def\na{\nabla}
\def\P{\Phi}
\begin{document}
\title{Enhancing the efficiency of open quantum batteries via adjusting the classical driving field}

\author{Maryam Hadipour}
\affiliation{Faculty of Physics, Urmia University of Technology, Urmia, Iran.}
\author{Soroush Haseli}
\email{soroush.haseli@uut.ac.ir}
\affiliation{Faculty of Physics, Urmia University of Technology, Urmia, Iran.}
\affiliation{
School of Physics, Institute for Research in Fundamental Sciences (IPM), Tehran P.O. Box 19395-5531, Iran}
\date{\today}
\def\be{\begin{equation}}
  \def\ee{\end{equation}}
\def\bea{\begin{eqnarray}}
\def\eea{\end{eqnarray}}
\def\f{\frac}
\def\n{\nonumber}
\def\l{\label}
\def\p{\phi}
\def\o{\over}
\def\R{\rho}
\def\pa{\partial}
\def\om{\omega}
\def\na{\nabla}
\def\P{\Phi}

\begin{abstract}
In the context of quantum information, a quantum battery refers to a system composed of quantum particles that can store and release energy in a way that is governed by the principles of quantum mechanics. The study of open quantum batteries is motivated by the fact that real-world quantum systems are almost never perfectly isolated from their environment. One important challenge in the study of open quantum batteries is to develop theoretical models that accurately capture the complex interactions between the battery and its environment. the goal of studying open quantum batteries is to develop practical methods for building and operating quantum devices that can store and release energy with high efficiency and reliability, even in the presence of environmental noise and other sources of decoherence. The charging process of open quantum batteries under the influence of dissipative environment will be studied. In this Work, the effect of the classical driving field on the charging process of open quantum batteries will be investigated.  The classical driving field can be used to manipulate the charging and discharging process of the battery, leading to enhanced performance and improved efficiency. It also will be showed that the efficiency of open quantum batteries depends on detuning between the qubit and the classical driving field and  central frequency of the cavity and the classical driving field.
\end{abstract}
\maketitle

\section{INTRODUCTION}	.
Quantum batteries are a recent development in quantum information science that have the potential to revolutionize energy storage technology. A quantum battery is a device that stores energy in the form of quantum states, and it can be charged and discharged using quantum mechanical processes. Compared to classical batteries, quantum batteries offer the potential for increased energy storage capacity, faster charging and discharging times, and improved efficiency. Another key advantage of quantum batteries is their potential for increased energy storage capacity. In a classical battery, energy is stored in the form of chemical reactions, which limit the amount of energy that can be stored. In contrast, quantum batteries can store energy in the form of coherent superpositions of quantum states, which can have much higher energy densities than classical systems. This increased energy density can lead to smaller and more compact batteries with higher energy storage capacity. Several theoretical and experimental studies have been conducted to explore the potential of quantum batteries \cite{a1,a2,a3,a4,a5,a6,a7,a8,a9,a10,a11,a12,a13,a14,a15,a16,a17,a18,a19,a20,a21,a22,a23,a24,a25,a26}. 
As mentioned, extensive research has been conducted on quantum batteries, especially on the use of quantum resources to achieve an optimal quantum battery. Such a QB should not only possess high charging efficiency but also be capable of delivering the maximum amount of stored energy to the point of consumption \cite{a9,a10,a11}. Initially, many researchers focused on quantum batteries as a closed system \cite{a12,a13,a14}. In this case, the battery and charger are not affected by the environment. The demonstration of the quantum advantage in the charging power of a Dicke quantum battery is illustrated by utilizing closed Dicke quantum battery and closed Rabi quantum battery as instances in Ref.\cite{a12}. In Ref.\cite{a13},  the authors demonstrated that a two-photon closed Dicke quantum battery exhibits improved performance in terms of charging times and average charging power when compared to the single-photon case, due to the effects of two-photon coupling. Research has also been conducted on a closed quantum battery that comprises a group of two-level atoms\cite{a14}. Studies have revealed that non-interacting two-level atoms can be completely charged using a harmonic charging field. Additionally, it has been reported that a closed quantum system composed of N independent two-level atoms can be charged through a time-dependent classical resource. The aforementioned research focuses on achieving the optimal quantum battery by exploring ways to accelerate the charging process of the closed quantum battery. As real-world systems interact with the environment \cite{a14p}, it is crucial to also investigate open quantum batteries \cite{a14,a15,a16,a17,a18,a19,a20,a21}. In recent studies \cite{a22,a23,a24,a25}, it has been demonstrated that by implementing appropriate design strategies, it is possible to significantly minimize the adverse impact of the environment on the functionality of quantum batteries. In some scenarios, the environment can actually enhance the efficacy of quantum batteries. As an example, the authors of a particular study \cite{a7} propose an open quantum battery protocol that employs dark states, which can lead to increased capacity and power density. In their method, non-interacting spins are coupled to a reservoir. In Refs. \cite{a25,a26}, the authors have demonstrated that strong couplings between the system and the environment can improve the charging capabilities of quantum batteries. However, in most instances, the coupling between the system and the environment is quite weak, which can significantly diminish the charging performance of the quantum battery. Thus, improving the performance of quantum batteries in scenarios where the system-environment coupling is weak is a critical issue. The focus of this work is to investigate the charging mechanism of a quantum battery when operating under the weak and strong system-environment coupling regime. The focus of our investigation is  on a two-qubit system that models a charger-battery setup. This model enables us to explore how the battery can be charged without a direct connection to the charger, using Markovian and non-Markovian dynamics in a wireless-like charging process. This research suggests a method in which an external classical field can drive the dynamics of QB and charger, each with different transition frequencies, within a cavity-based system. QB and charger interact with the quantized modes of a high-Q cavity and are also coupled with the external classical field. The impact of the classical driving field on the QB behavior will be analyzed. The environment effectively charges the battery when it's in the strong coupling regime, however in the weak coupling regime the other parameter can control the charging process of QB . It can be shown that QB can be charge in both strong an weak coupling regimes. we examine the scenario where QB and charger have identical transition frequencies. It will be showed that the classical deriving field can enhance the performance of an open quantum battery i.e., the energy stored,
average charging power, and the extractable work. In conclusion, we examine how detuning impacts the performance of open quantum battery.  This paper is organized as follows. Sec. \ref{sec2} provides an overview of the energy stored in a quantum battery, the average power required for charging, and the amount of work that can be extracted from it. In Sec.\ref{sec3}, we introduce our physical model and derive the precise equations that govern the temporal behavior of the QB system, which is coupled to environment. .Section \ref{sec4} provides a summary of the conclusion.

\section{Quantm battery: Energy, Avarage power, Ergotropy}\label{sec2}
For a quantum battery to be considered good, it must fulfill two criteria: firstly, it should be able to store the maximum amount of energy in the shortest possible time; secondly, it should have the capability to discharge the stored energy sufficiently within the required time. In order to develop an effective quantum battery, it is necessary to analyze the performance of the battery itself, which includes metrics such as the amount of energy it can store, the average power required for charging, and the amount of work that can be extracted from it.  The energy stored in the quantum battery at a given time $t$ is defined as
\begin{equation}\label{ene}
E_B=\operatorname{Tr}\left[H_B \rho_B(t)\right]-\operatorname{Tr}\left[H_B \rho_B(0)\right],
\end{equation}
where $H_B$ is the Hamiltonian of QB, $\rho_B(t)$ is the  state of quantum battery at time $t$. The average power required for charging the quantum battery at time $t$ is given by $P_B=\frac{E_B}{t}$. In addition, the concept of ergotropy has been introduced in the field of quantum thermodynamics to represent the highest amount of work that can be obtained from a quantum battery once it has been charged using specific cyclic unitary operations.
\begin{equation}\label{org}
W_B=\operatorname{Tr}\left(\rho_B(t) H_B\right)-\operatorname{Tr}\left(\sigma_{\rho_B} H_B\right)
\end{equation}
where $\sigma_{\rho_B}$ is called passive state. If the state of the quantum battery is a passive state, then it is not possible to extract any work from it using cyclic unitary operations.  The density matrix of the quantum battery and its associated Hamiltonian can be expressed using spectral decomposition as 
\begin{eqnarray}
\rho_B&=&\sum_{i=1}^{d} r_i \vert r_i \rangle \langle r_i \vert \quad r_1 \geq r_2 \geq ... \geq r_d, \\
H_B&=&\sum_{i=1}^{d} \varepsilon_i \vert \varepsilon_i \rangle \langle \varepsilon_i \vert \quad \varepsilon_1 \leq \varepsilon_2 \leq ... \leq \varepsilon_d,
\end{eqnarray}
n the above equation, $d$ denotes the dimension of the Hilbert space, while $r_i$($\varepsilon_i$) and  $\vert r_i \rangle $($\vert \varepsilon_i \rangle$) represent the eigenvalues (eigenstates) of the density matrix $\rho_B$ and the Hamiltonian $H_B$, respectively. The passive state $\sigma_{\rho_B}$  is characterized by a non-increasing probability distribution with respect to its associated Hamiltonian $H_B$ as $\sigma_{\rho_B}=\sum_{i=1}^{d} r_i \vert \varepsilon_i \rangle \langle \varepsilon_i \vert$, and the commutator between $H_B$ and $\sigma_{\rho_B}$ is zero, i.e., $\left[ H_B,\sigma_{\rho_B}\right]=0$. So, from Eq/\ref{org}, the ergotropy can be obtained as
\begin{equation}
W_B=\sum_{i,j}^d r_i \varepsilon_j\left(\left|\left\langle r_i \mid \varepsilon_j\right\rangle\right|^2-\delta_{i, j}\right),
\end{equation}  
where $\delta_{i, j}$ is the Kronecker delta function. In order to evaluate the the most efficient quantum battery, the highest possible values for internal energy $E_{B,\max}$, power output $P_{B,\max}$, and ergotropy $W_{B,\max}$ need to be determined
\begin{equation}
\begin{aligned}
E_{B,\max} & =\max _t\left[E_B(t)\right]=E_B\left(t_E\right) \\
P_{B,\max} & =\max _t\left[P_B(t)\right]=P_B\left(t_P\right) \\
W_{B,\max} & =\max _t\left[W_B(t)\right]=W_B\left(t_W\right)
\end{aligned}
\end{equation}
Where the values $t_E$, $t_P$, and $t_W$ refer to the time at which the maximum internal energy $E_{B,\max}$, maximum power output $P_{B,\max}$, and maximum ergotropy $W_{B,\max}$ of a quantum battery are achieved, respectively. When analyzing the charging process of a quantum battery, it is desirable to have larger values of $E_B$ (maximum internal energy), $P_B$ (maximum power output), and $W_B$ (maximum ergotropy), as demonstrated in the subsequent sections. 

\section{The model}\label{sec3}
\begin{figure}[H]
\centering
    \includegraphics[width =1 \linewidth]{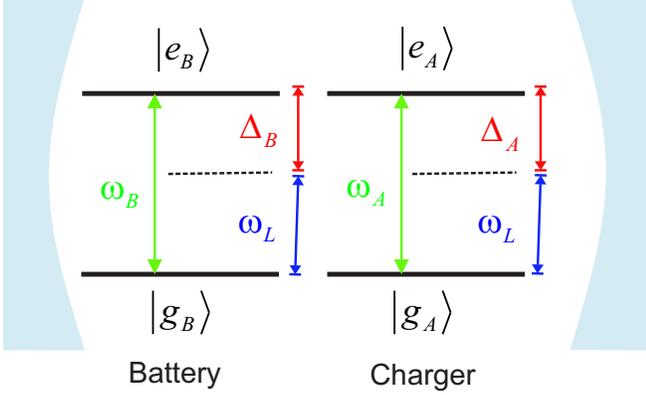}
    \caption{A visual representation of the model where a classical field is used to drive the quantum battery within a leaky cavity }
    \label{Fig1}
  \end{figure}

The setup is considered where two QB and charger with distinct transition frequencies $\omega_j$ (where $j = 1$ and $2$) are subjected to an external classical field. Additionally, we make the assumption that the QB and charger are interacting with a common zero-temperature environment composed of the quantized modes of a high-Q cavity, as depicted in Fig.\ref{Fig1}. The Hamiltonian that characterizes the entire system, under the assumptions of the dipole and rotating wave approximations, can be expressed as

\begin{eqnarray}\label{H1}
\hat{H}&=&\sum_{j=A}^B \frac{\omega_j}{2} \hat{\sigma}_z^{(j)}+\sum_k \omega_k  \hat{a}_k^{\dagger} \hat{a}_k \nonumber \\
&+&\sum_{j=A}^B\left(\sum_k \alpha_j g_k \hat{a}_k \hat{\sigma}_{+}^{(j)}+\Omega e^{-i \omega_L t} \hat{\sigma}_{+}^{(j)}+\text { H.c. }\right)
\end{eqnarray}
where $\sigma_z^{j}$ is the Pauli operator of $j$th qubit with transition frequency $\omega_j$, $\omega_L$ and $\omega_k$ stand for the frequencies of the classical driving field and the quantized modes of the cavity, respectively. $\hat{\sigma}_{+}^{(j)}$ and $\hat{\sigma}_{-}^{(j)}$  represents the operator that raises (or lowers) the state of the $j$th qubit, respectively.  $\hat{a}_k$ and ($\hat{a}_k^{\dagger}$) are the annihilation (and creation) operators for the $k$th mode of the cavity. Furthermore, $\Omega$ and $g_k$ denote the strengths of the couplings between the qubits and the classical driving field and between the cavity modes and the qubits, respectively. Additionally, the parameter $\alpha_j$ quantifies the interaction  of the $j$th qubit with its surrounding environment, and it has no specific unit. We assume that $\Omega$ is a real number, and its value is negligible in compared with the frequencies of the atom and laser, that is, the value of $\Omega$ is significantly smaller than $\Omega_j$ and $\Omega_L$.

The eigenvalues of the system  remain unaffected by a unitary transformation \cite{a27,a28}. As a result, we take into consideration the unitary transformation $U=e^{-i \omega_L\left(\hat{\sigma}_z(A)+\hat{\sigma}_z(B) t / 2\right.}$, which enables us to express the Hamiltonian of the system in the rotating reference frame as
\begin{equation}\label{h2}
\begin{aligned}
\hat{H}_{\mathrm{eff}}&=\hat{H}_{\mathrm{1}}+\hat{H}_{\mathrm{2}},\\
\hat{H}_{\mathrm{1}}&=\left(\frac{\Delta_A}{2} \hat{\sigma}_z^{(A)}+\frac{\Delta_B}{2} \hat{\sigma}_z^{(B)}\right)+\Omega\left(\hat{\sigma}_x^{(A)}+\hat{\sigma}_x^{(B)}\right), \\
\hat{H}_{\mathrm{2}} & =\sum_k \omega_k \hat{a}_k^{\dagger} \hat{a}_k+\sum_{j=A}^B\left(\sum_k \alpha_j g_k \hat{a}_k \hat{\sigma}_{+}^{(j)} e^{i \omega_L t}+\text { H.c. }\right),
\end{aligned}
\end{equation}
Where, $\Delta_j=\omega_j - \omega_L$, denoted the difference or detuning between transition frequency of qubit and the frequency of classical driving field. Upon analysis of Eq.\ref{h2}, it becomes clear that the Hamiltonian operator $\hat{H}_1$ can be represented as $\hat{H}_1=\hat{H}_1^{A}+\hat{H}_2^{B}$, where $\hat{H}_1^{j}=\frac{\Delta_j}{2} \hat{\sigma}_z^{(j)}+\Omega \hat{\sigma}_x^{(j)}$, with $j=A,B$. Eigenstates  of $\hat{H}_1^{j}$ can be obtain as  
\begin{eqnarray}\label{dress}
|E\rangle_j&=&\sin \frac{\eta_j}{2}|g\rangle_j+\cos \frac{\eta_j}{2}|e\rangle_j, \nonumber \\
|G\rangle_j&=&\cos \frac{\eta_j}{2}|g\rangle_j-\sin \frac{\eta_j}{2}|e\rangle_j
\end{eqnarray}
where $|E\rangle_j$ and $|G\rangle_j$ are called dressed bases and $\eta_j=\operatorname{Arctan}\left[2 \Omega / \Delta_j\right]$. In this bases, the effective Hamiltonian can be reformulated as 
\begin{eqnarray}
\hat{H}_{\mathrm{eff}}&=&\frac{\chi_A}{2} \hat{\varrho}_z^{(A)}+\frac{\chi_B}{2} \hat{\varrho}_z^{(B)}+\sum_k \omega_k \hat{a}_k^{\dagger} \hat{a}_k \nonumber \\
&+&\left(\alpha_A \cos ^2\left(\eta_A / 2\right) \hat{\varrho}_{+}^{(A)}+\alpha_B \cos ^2\left(\eta_B / 2\right) \hat{\varrho}_{+}^{(B)}\right)\times \nonumber \\
&\times& e^{+i \omega_L t} \sum_k g_k \hat{a}_k+\text { H.c. }
\end{eqnarray}
where $\hat{\varrho}_z^{(j)}=|E\rangle_j\langle E|-| G\rangle_j\langle G|$, $\chi_j=\sqrt{\Delta_j^2+4 \Omega^2}$ and $\hat{\varrho}_{+}^{(j)}=|E\rangle_j\langle G|\left(\hat{\varrho}_{-}^{(j)}=|G\rangle_j\langle E|\right)$ is the lowering (rising) operator in dressed bases. It is important to mention that while obtaining the effective Hamiltonian (4), the energy terms that do not conserve energy were ignored using the rotating-wave approximation \cite{a29,a30}. Here it is necessary to introduce the  the collective coupling constant $\alpha_T=\left(\alpha_A^2+\alpha_B^2\right)^{1 / 2}$ and relative parameters $r_j=\alpha_j/\alpha_T$. The  relative parameters should  satisfy the condition $r_A^2+r_B^2=1$. In addition, by changing $\alpha_T$, the strong and weak coupling regimes can be explored. At this point, it is possible to investigate how the classical field affects the performance  of the quantum battery. We assume that in the new dressed states, the initial state of the entire system is given by
\begin{equation}\label{ini}
\left|\psi_0\right\rangle=\left(c_{01}|E\rangle|G\rangle+c_{02}|G\rangle|E\rangle\right) \otimes|\mathbf{0}\rangle_R,
\end{equation}
where $|\mathbf{0}\rangle_R$ is the molti-mode vacuum state of the leaky cavity. The initial state of entire system in Eq.\ref{ini} can be transformed into the following state 
\begin{eqnarray}\label{fin}
|\psi(t)\rangle&=&C_A(t)|E\rangle|G\rangle|\mathbf{0}\rangle_R+C_B(t)|G\rangle|E\rangle|\mathbf{0}\rangle_R \nonumber \\
&+&\sum_k C_k(t)|G\rangle|G\rangle\left|\mathbf{1}_k\right\rangle
\end{eqnarray}
in which the $\left|\mathbf{1}_k\right\rangle$ denotes the molti-mode state  of the leaky cavity with one photon in $k$th mode and vacuum in other modes.  By using Eq.\ref{fin} and putting it in the Schrodinger equation, we get the following integro-differential
equations for amplitudes coefficients $C_1(t)$, $C_2(t)$ and $C_k(t)$
\begin{equation}\label{integ1}
\dot{C}_j(t)=-i \alpha_j \cos ^2\left(\eta_j / 2\right) e^{i \omega_L t} \sum_k g_k C_k(t) e^{-i\left(\omega_k-\chi_j\right) t}
\end{equation}

\begin{eqnarray}\label{integ2}
    \begin{split}
\dot{C}_k(t)=-i g_k^* e^{-i \omega_L t}\left(\alpha_1 \cos ^2\left(\eta_1 / 2\right) e^{i\left(\omega_k-\chi_1\right) t} C_1(t) \right.  \\
 \left. +\alpha_2 \cos ^2\left(\eta_2 / 2\right) e^{i\left(\omega_k-\chi_2\right) t} C_2(t)\right)
\end{split}
\end{eqnarray}
By integrating  Eq.\ref{integ2} and substituting the result into Eq.\ref{integ1}, two integro-differential equations for $C_1(t)$ and $C_2(t)$ can be obtained as
\begin{eqnarray}\label{c1}
    \begin{split}
\dot{C}_1(t)=-\int_0^t \mathrm{~d} t^{\prime} G\left(t-t^{\prime}\right)\left(\alpha_A^2 \cos ^4\left(\eta_A / 2\right) C_1\left(t^{\prime}\right)+ \right.  \\
 \left. +\alpha_A \alpha_B \cos ^2\left(\eta_A / 2\right) \cos ^2\left(\eta_2 / 2\right) e^{-i \xi_{BA} t^{\prime}} C_2\left(t^{\prime}\right)\right) e^{i \xi_A\left(t-t^{\prime}\right)}
\end{split}
\end{eqnarray}

\begin{eqnarray}\label{c2}
    \begin{split}
\dot{C}_2(t)=-\int_0^t \mathrm{~d} t^{\prime} G\left(t-t^{\prime}\right)\left(\alpha_A \alpha_B \cos ^2\left(\eta_A / 2\right) \cos ^2\left(\eta_B / 2\right)\times \right.  \\
 \left.\times e^{i \xi_{BA} t^{\prime}} C_1\left(t^{\prime} \right) + \alpha_B^2 \cos ^4\left(\eta_B / 2\right) C_2\left(t^{\prime}\right)\right) e^{i \xi_B\left(t-t^{\prime}\right)},
\end{split}
\end{eqnarray}
where the correlation function, represented by the kernel $G\left(t-t^{\prime}\right)$, is defined using the continuous limits of the environment frequency as
\begin{equation}\label{cf}
G\left(t-t^{\prime}\right)=\int \mathrm{d} \omega J(\omega) e^{i\left(\omega_L+\omega_c-\omega\right)\left(t-t^{\prime}\right)}
\end{equation}
Here the Lorentzian spectral density $J(\omega)=W^2 \lambda / \pi\left[\left(\omega-\omega_c\right)^2+\lambda^2\right]$ is considered.  In this context, the variable W varies in proportion to the vacuum  Rabi frequency as $\mathcal{R}=\alpha_T W$, while the variable $\lambda$ denotes the rate of cavity losses. $\omega_c$ is the frequency of the cavity, $\xi_j=\chi_j-\omega_c$ and $\xi_{BA}=\xi_B-\xi_A$. The set of coupled differential equations mentioned above can theoretically be solved analytically using the Laplace transformation method. Nevertheless, it will be explained that a straightforward analytical solution for the amplitude coefficients can only be obtained in the specific scenario where  $\omega_A$ equals to $\omega_B$.
\section{Similar quantum battery and charger}

In this section, we will discuss a specific scenario where quantum battery and charger  have identical transition frequencies $\omega_A=\omega_B=\omega_0$. In this case we have $\chi_A=\chi_B \equiv \chi, \Delta_A=\Delta_B \equiv \Delta, \eta_A=\eta_B \equiv \eta$ and $\xi_A=\xi_B \equiv \xi$. By this consideration the two integro-differential
equations in Eqs. \ref{c1} and \ref{c2}, reduce to following equation 
\begin{eqnarray}\label{c3}
\dot{C}_i(t)&=&-\cos ^4(\eta / 2) \times \nonumber \\
&\times& \int_0^t \mathrm{~d} t^{\prime} F\left(t-t^{\prime}\right)\left(\alpha_i^2 C_i\left(t^{\prime}\right)+\alpha_i \alpha_j C_j\left(t^{\prime}\right)\right),
\end{eqnarray}
where $i \neq j$ and $(i,j) \in \left\lbrace A,B\right\rbrace $.  Before analyzing the performance of QB under classical driving field, it's worthwhile to initially look for a stationary solution for above integro-differential equations in Eq.\ref{c3}, where $C_j(t \rightarrow \infty)$ remains constant. One way to achieve this is by assigning $\dot{C}_j(t)=0$ in Eq.\ref{c3}. This will result in the creation of a long-lasting state that is free from decoherence, which can then be normalized as
\begin{equation}
\left|\phi_{-}\right\rangle=r_2|E\rangle|G\rangle-r_1|G\rangle|E\rangle .
\end{equation}
It can be observed that the presence of sub-radiant state's  is not contingent on factors such as the classical driving field, the type of spectral density, or whether the system is in resonance or off-resonance. The initial state of the entire system comprises two components: first the sub-radiant state $\left|\phi_{-}\right\rangle$ and the second super-radiant state $\left|\phi_{+}\right\rangle$. In contrast to the sub-radiant state, the super-radiant state changes over time with the survival amplitude $\mathcal{Z}(t)$, which satisfy the following equation of motion
\begin{equation}\label{zar}
\dot{\mathcal{Z}}(t)=-\alpha_T^2 \cos ^4(\eta / 2) \int_0^t F\left(t-t^{\prime}\right) \mathcal{Z}\left(t^{\prime}\right) \mathrm{d} t^{\prime},
\end{equation}
Suppose that the analytical form of $\mathcal{Z}(t)$ is derived. In that case, it becomes effortless to demonstrate that the amplitudes $C_i(t)$ can be written as
\begin{eqnarray}\label{dyn}
C_1(t)&=&r_2 \beta_{-}+r_1 \mathcal{Z}(t) \beta_{+}, \nonumber \\
C_2(t)&=&-r_1 \beta_{-}+r_2 \mathcal{Z}(t) \beta_{+},
\end{eqnarray}
Here, $\beta_{\pm}$ have been defines as $\beta_{ \pm} \equiv\left\langle\psi_{ \pm} \mid \psi(0)\right\rangle$. From Eq.\ref{dyn}, It can be seen that the dynamics of quantum battery and charger depends on $\mathcal{Z}(t)$. From Eq.\ref{cf}, the correlation function $F(t-t^{\prime})$ can be obtained as 
\begin{equation}
F\left(t-t^{\prime}\right)=W^2 e^{-\lambda\left(t-t^{\prime}\right)} e^{i\left(\chi+\Delta_L\right)\left(t-t^{\prime}\right)}
\end{equation}
where $\Delta_L$ is the detuning between the classical driving field $\omega_L$ and central frequency of the cavity $\omega_c$, i.e. $\Delta_L=\omega_L-\omega_c$. Using Laplace transform method from Eq.\ref{zar}, $\mathcal{Z}(t)$ can be obtained as
\begin{equation}
\mathcal{Z}(t)=e^{-M t / 2}\left(\cosh (\mathcal{F} t / 2)+\frac{M}{\mathcal{F}} \sinh (\mathcal{F} t / 2)\right)
\end{equation} 
where $M=\lambda-i\left(\chi+\Delta_L\right)$ and $\mathcal{F}=\sqrt{M^2-\alpha_T^2 W^2(1+\cos \eta)^2}$. To determine weak and strong coupling regimes in this model, parameter $R$ is defined as $R=\frac{\mathcal{R}}{\lambda}$, such that if $R$ greater than one, the coupling is strong and if it is smaller than one, it is weak.

By using Eq. \ref{fin}, it is possible to express the reduced density operator of both the QB and the charger at a time $t=\tau$ as
\begin{equation}\label{den}
\begin{aligned}
& \rho_{B}(\tau)= \vert C_2(\tau)\vert^{2}\vert e \rangle \langle e \vert_{B}+ \left( 1- \vert C_2(\tau)\vert^{2} \right)\vert g \rangle \langle g \vert_{B},  \\
& \rho_{\mathrm{A}}(\tau)=\vert C_1(\tau)\vert^{2}\vert e \rangle \langle e \vert_{A}+ \left( 1- \vert C_1(\tau)\vert^{2} \right)\vert g \rangle \langle g \vert_{A}.
\end{aligned}
\end{equation}
Using Eqs.\ref{ene} and \ref{den} , the stored energy in quantum battery can be obtained as $E_B= \vert C_2(\tau) \vert^2  \chi$. 
Here, it is assumed that the quantum battery is starting with no stored energy  and charger has maximum energy. Based on assumption the initial state of entire system should be $|\psi_0\rangle=|e\rangle_A|g\rangle_B \otimes|0\rangle_{\mathrm{R}}$(in Eq. \ref{ini}, set $c_{01}=1$ and $c_{02}=0$). The average charging power at time $\tau$ is given by $P_B=\vert C_2(\tau) \vert^2  \chi /\tau$. From Eqs. \ref{org} and \ref{den}, the ergotropy can be obtained as
\begin{equation}
W_B=\left(2 \vert C_2(t) \vert^{2} -1 \right) \Theta(C_2(t) \vert^{2} -\frac{1}{2})\chi,
\end{equation} 
where $\Theta(...)$ is Heaviside function. The maximum storage energy, storage power and ergotropy  can be obtained 
\begin{equation}
\begin{aligned}
E_{B,\max} & =\max_\tau\left[\vert C_2(\tau) \vert^2  \chi\right],  \\
P_{B,\max} & =\max_\tau\left[\vert C_2(\tau) \vert^2  \chi /\tau\right] , \\
W_{B,\max} & =\max_\tau\left[\left(2 \vert C_2(t) \vert^{2} -1 \right) \Theta(C_2(t) \vert^{2} -\frac{1}{2})\chi\right]. 
\end{aligned}
\end{equation}
\begin{figure}\label{Fig2}
\centering
    \includegraphics[width =1 \linewidth]{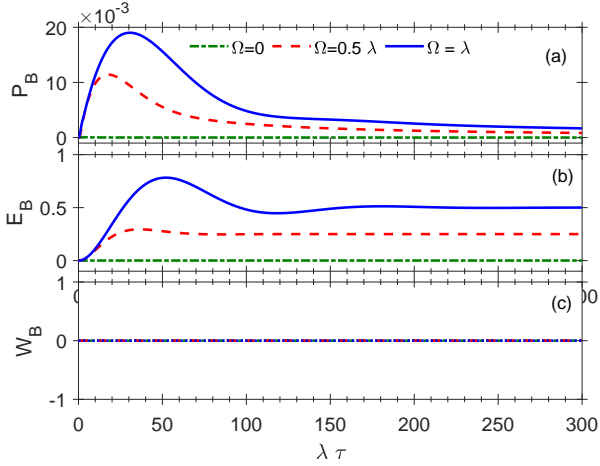}
      \vspace*{-15mm}
    \caption{(a)The charging power $P_B$ (b)The internal energy $E_B$ and (c) Ergotropy $W_B$, as a function of the charging time $\tau$ in weak coupling regime with $R=0.5$ for different values of the coupling strength between QB and classical driving field $\Omega$ in resonance case $\Delta=\Delta_L=0$ and $r_1=\frac{1}{\sqrt{2}}$. }
    \label{Fig2}
  \end{figure}
\subsection{Weak coupling regime}
In this section, the results will be shown for weak coupling regime . Here we consider the case in which $R=0.5$. 
In Fig.\ref{Fig2}, the performance of the quantum battery has represented as a function of $\lambda \tau$ in weak coupling regime with $R=0.5$ for resonance case $\Delta=\Delta_L=0$. In
these plots, we have set $r_1=\frac{1}{\sqrt{2}}$. Fig.\ref{Fig2}(b), shows the stored energy in quantum battery as a function of $\lambda \tau$ in strong coupling regime. As can be seen the stored energy increases with increasing the strength of the classical deriving field. As can be seen in the absence of the classical driving field the charging power is weak and near to zero while with the appearance of the classical driving field and increasing its strength $\Omega$, the amount of charging power $P_B$ increases. In Fig.\ref{Fig2}(c), the ergotropy is plotted in terms of $\lambda \tau$. It can be see that in weak coupling regime, even with the presence of a classical field with resonance, the ergotropy value is zero at all times. As a summary, in this case, the classical field has an effective and positive role in stored energy and charging power of quantum battery, while it has no effect on improving the work that can be extracted from the battery.
\begin{figure}\label{Fig3}
\centering
    \includegraphics[width =1 \linewidth]{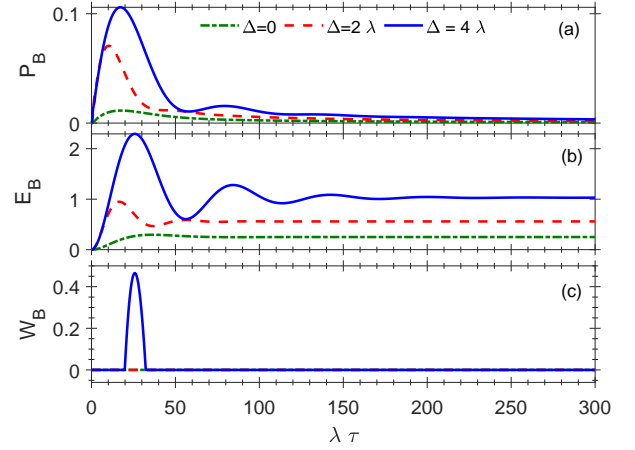}
      \vspace*{-15mm}
    \caption{(a)The charging power $P_B$ (b) The internal energy $E_B$ and (c) Ergotropy $W_B$, as a function of the charging time $\tau$ in weak coupling regime with $R=0.5$ for different values of the detuning between the qubit and the classical driving field $\Delta$ with $\Delta_L=0$ and $r_1=\frac{1}{\sqrt{2}}$. }
    \label{Fig3}
  \end{figure}
Fig.\ref{Fig3} shows the effect of detuning between the qubit and the classical driving field $\Delta$ on the performance of the quantum battery in weak coupling regime $R=0.5$. In Fig.\ref{Fig3}(a), the charging power has plotted in terms of $\lambda \tau$. As can be seen increasing the detuning $\Delta$ leads to the enhancement of charging power of quantum battery. Increasing the detuning $\Delta$ also improves the stored energy in the quantum battery, this fact is shown in Fig.\ref{Fig3}(b).It can be seen that in weak coupling regime the charging power increases with increasing $\Delta$. Although in this subsection the weak coupling regime has been considered, while in contrast to the resonace case, it is observed from Fig.\ref{Fig3}(c) that for large values of detuning $\Delta$ the work that can be extracted from the quantum battery is non-zero.
\begin{figure}\label{Fig4}
\centering
    \includegraphics[width =1 \linewidth]{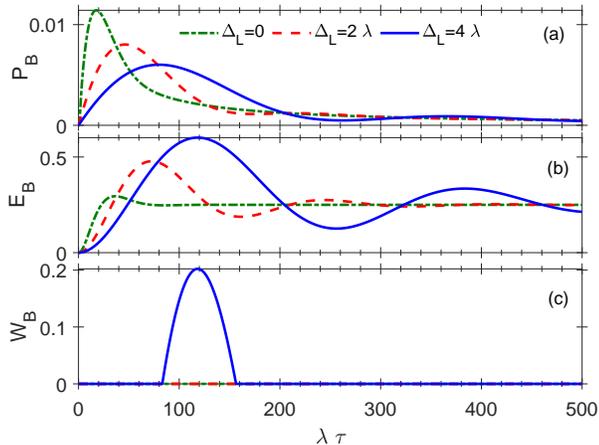}
      \vspace*{-15mm}
    \caption{(a)The  charging power $P_B$ (b)The internal energy $E_B$  and (c) Ergotropy $W_B$, as a function of the charging time $\tau$ in weak coupling regime with $R=0.5$ for different values of $\Delta_L$ (the detuning between the classical driving field  and central frequency of
the cavity ), $\Delta=0$ and $r_1=\frac{1}{\sqrt{2}}$. }
    \label{Fig4}
  \end{figure}  
In Fig.\ref{Fig4}, the effect of the detuning between the classical driving field  and central frequency of the cavity $\Delta_L$ on the performance of QB has been represented. From Fig.\ref{Fig4}(a), it can be seen that increasing $\Delta_L$ has a destructive effect on the charging power of the quantum battery. This means that increasing $\Delta_L$ will decrease the charging power of the quantum battery. 
When the frequency of classical driving field $\omega_L$ is detuned from the central frequency of the cavity $\omega_c$, it means that the frequency of the driving field is not exactly matched to the resonant frequency of the cavity. As a result, the energy transfer between the driving field and the cavity mode is less efficient. This is because the driving field and the cavity mode are not in resonance, and there is a phase difference between them. The efficiency of energy transfer is further reduced when the detuning is increased. This is because the cavity mode becomes less sensitive to the driving field as the detuning increases. This reduces the effective coupling between the driving field and the cavity mode, and hence reduces the amount of energy that can be transferred to the battery. So, increasing the detuning between the classical driving field and the central frequency of the cavity decreases the charging power of the quantum battery in a leaky cavity.
The situation is different for stored energy of quantum battery. As can be seen from Fig.\ref{Fig4}(b), the stored energy increases with increasing the detuning $\Delta_L$. Fig. \ref{Fig4}(c), shows the effect of $\Delta_L$ on ergotropy. As can be seen, the ergotropy has non-zero value for large value of $\Delta_L$.  
\begin{figure}\label{Fig5}
\centering
    \includegraphics[width =1 \linewidth]{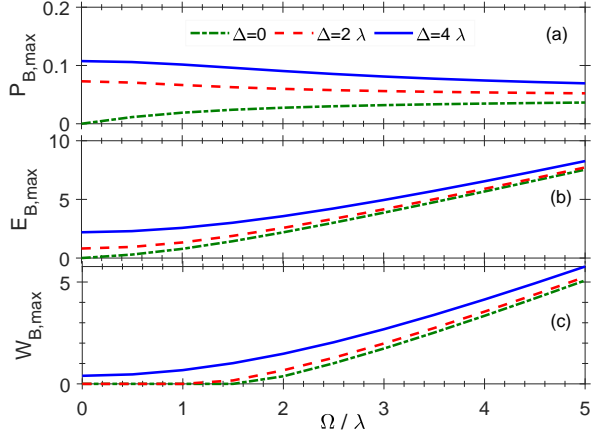}
      \vspace*{-15mm}
    \caption{(a)The maximum charging power $P_{B,\max}$ (b)The maximum internal energy $E_{B,\max}$(c)The maximum Ergotropy $W_{B,\max}$, as a function of the  classical deriving field strength $\Omega$ in weak coupling regime with $R=0.5$ for different values of $\Delta$ , $\Delta_L=0$ and $r_1=\frac{1}{\sqrt{2}}$. }
    \label{Fig5}
  \end{figure}
In Fig.\ref{Fig5}(a), the maximum charging power has been plotted as a function of   the  classical deriving field strength $\Omega$ for different value of $\Delta$ when $\Delta_L=0$. As can be seen for resonance case $\Delta=Delta_L=0$, the result is consistent with the results of Fig.\ref{Fig2}(a), i.e. the maximum charging power $P_{B,\max}$ increases with increasing the $\Omega$. While, the situation is different for $\Delta \neq 0$ and the charging power decreasing with increasing the strength of the classical deriving field, while, the maximum charging power increases with increasing $\Delta$.  When there is a detuning between the frequency of the classical driving field and the transition frequency of the quantum system, the absorption of photons becomes less efficient. This is because the energy difference between the initial and final states of the quantum system is not exactly equal to the energy of the photons in the driving field. As a result, the probability of absorption decreases, which in turn decreases the charging power of the quantum battery. Fig.\ref{Fig5}(b), represents the maximum stored energy in terms of $\Omega$. As can be seen the maximum stored energy  increases with increasing the $\Omega$ for different values of $\Delta$. It can also be seen that the maximum stored energy increases with increasing $\Delta$. Fig.\ref{Fig5}(c), shows that the ergotropy increases with increasing the $\Omega$ for resonant and non-resonant case. 
  \begin{figure}\label{Fig6}
\centering
    \includegraphics[width =1 \linewidth]{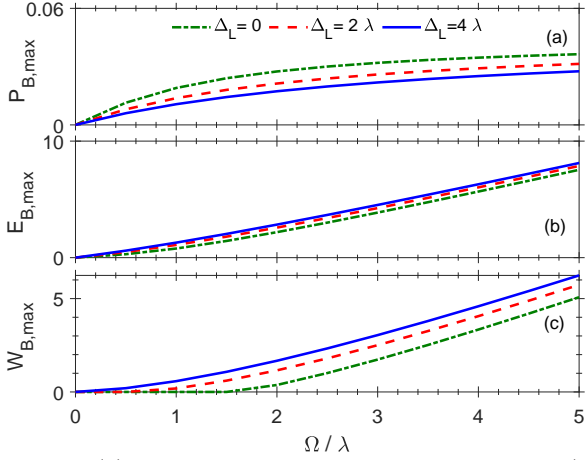}
      \vspace*{-15mm}
    \caption{(a)The maximum charging power $P_{B,\max}$ (b)The maximum internal energy $E_{B,\max}$(c)The maximum Ergotropy $W_{B,\max}$, as a function of the charging time $\tau$ in weak coupling regime with $R=0.5$ for different values of $\Delta_L$ (the detuning between the classical driving field  and central frequency of
the cavity ), $\Delta=0$ and $r_1=\frac{1}{\sqrt{2}}$. }
    \label{Fig6}
  \end{figure}
In Fig.\ref{Fig6}(a), the maximum charging power has been plotted as a function of   the  classical deriving field strength $\Omega$ for different value of $\Delta_L$ when $\Delta=0$. As can be seen for different value of $\Delta_L$ the maximum charging power increases with increasing the $\Omega$. It can also be seen that increasing $\Delta_L$ decreases the maximum charging power. The maximum stored energy in quantum battery has been plotted in terms of $\Omega$ for different values of $\Delta_L$ . It can be seen that the maximum stored energy increases with increasing $\Omega$ and $\Delta_L$. 
\subsection{Strong coupling regime}
In this section, the results will be shown for weak coupling regime . Here we consider the case in which $R=10$.
\begin{figure}\label{Fig7}
\centering
    \includegraphics[width =1 \linewidth]{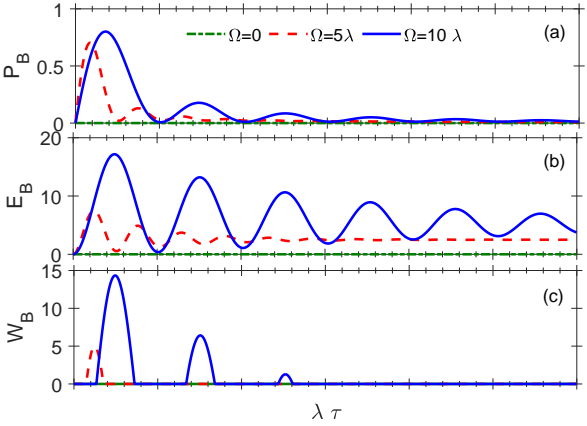}
      \vspace*{-15mm}
    \caption{(a)The charging power $P_B$ (b)The internal energy $E_B$ and (c) Ergotropy $W_B$, as a function of the charging time $\tau$ in strong coupling regime with $R=10$ for different values of the coupling strength between QB and classical driving field $\Omega$ in resonance case $\Delta=\Delta_L=0$ and $r_1=\frac{1}{\sqrt{2}}$. }
    \label{Fig7}
  \end{figure}
 In Fig.\ref{Fig7}, the performance of the quantum battery has represented as a function of $\lambda \tau$ in strong coupling regime with $R=10$ for resonance case $\Delta=\Delta_L=0$. In
these plots, we have set $r_1=\frac{1}{\sqrt{2}}$. Fig.\ref{Fig7}(a), shows the charging power $P_B$ as a function of $\lambda \tau$ for different strength of classical driving field  $\Omega$. As can be seen in the absence of the classical driving field the charging power is weak and near to zero while with the appearance of the classical driving field and increasing its strength $\Omega$, the amount of charging power $P_B$ increases. Fig.\ref{Fig7}(b), shows the stored energy in quantum battery as a function of $\lambda \tau$ in strong coupling regime. As can be seen the stored energy increases with increasing the strength of the classical deriving field. In Fig.\ref{Fig7}(c), the ergotropy is plotted in terms of $\lambda \tau$. It can be see that in strong coupling regime, the ergotropy value is non-zero, even in the absence of a classical field.
\begin{figure}\label{Fig8}
\centering
    \includegraphics[width =1 \linewidth]{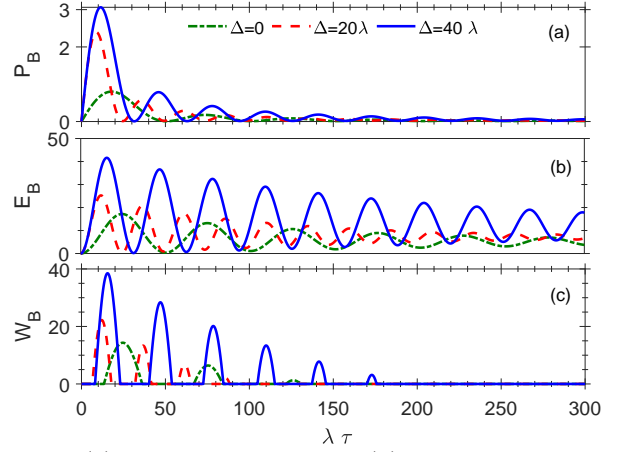}
      \vspace*{-15mm}
    \caption{(a)The charging power $P_B$ (b)The internal energy $E_B$ and (c) Ergotropy $W_B$, as a function of the charging time $\tau$ in strong coupling regime with $R=10$ for different values of the detuning between the qubit and the classical driving field $\Delta$ with $\Delta_L=0$ and $r_1=\frac{1}{\sqrt{2}}$. }
    \label{Fig8}
  \end{figure}
Fig.\ref{Fig8} shows the effect of detuning between the qubit and the classical driving field $\Delta$ on the performance of the quantum battery in strong coupling regime $R=10$. In Fig.\ref{Fig8}(a), the charging power has plotted in terms of $\lambda \tau$. As can be seen increasing the detuning $\Delta$ leads to the enhancement of charging power of quantum battery. Increasing the detuning $\Delta$ also improves the stored energy in the quantum battery, this fact is shown in Fig.\ref{Fig8}(b).It can be seen that in strong coupling regime the charging power increases with increasing $\Delta$. Although in this subsection the weak coupling regime has been considered, while in contrast to the resonace case, it is observed from Fig.\ref{Fig8}(c) that for large values of detuning $\Delta$ the work that can be extracted from the quantum battery is non-zero.
\begin{figure}\label{Fig9}
\centering
    \includegraphics[width =1 \linewidth]{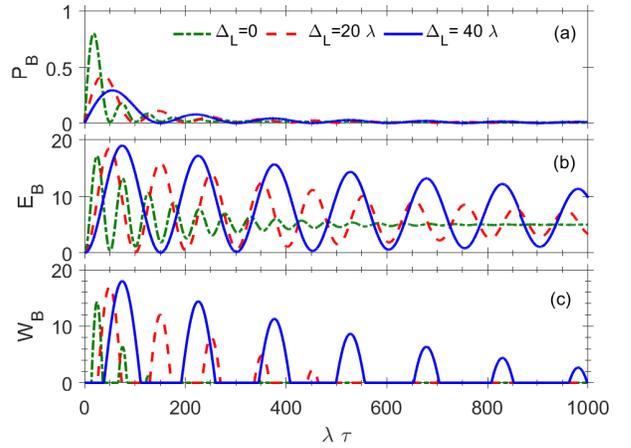}
      \vspace*{-15mm}
    \caption{(a)The charging power $P_B$ (b)The internal energy $E_B$ and (c) Ergotropy $W_B$, as a function of the charging time $\tau$ in strong coupling regime with $R=10$ for different values of $\Delta_L$ (the detuning between the classical driving field  and central frequency of
the cavity ), $\Delta=0$ and $r_1=\frac{1}{\sqrt{2}}$. }
    \label{Fig9}
  \end{figure}  

  \begin{figure}\label{Fig10}
\centering
    \includegraphics[width =1 \linewidth]{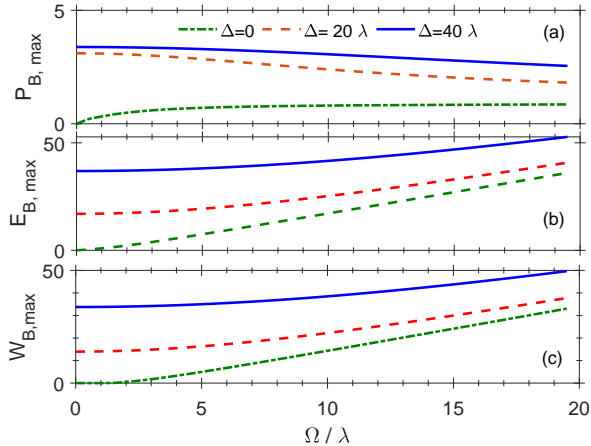}
      \vspace*{-15mm}
    \caption{(a)The maximum charging power $P_{B,\max}$ (b)The maximum internal energy $E_{B,\max}$(c)The maximum Ergotropy $W_{B,\max}$, as a function of the  classical deriving field strength $\Omega$ in strong coupling regime with $R=10$ for different values of $\Delta$ , $\Delta_L=0$ and $r_1=\frac{1}{\sqrt{2}}$. }
    \label{Fig10}
  \end{figure}
In Fig.\ref{Fig10}(a), the maximum charging power has been plotted as a function of   the  classical deriving field strength $\Omega$ for different value of $\Delta$ when $\Delta_L=0$ in strong coupling regime. As can be seen for resonance case $\Delta=Delta_L=0$, the result is consistent with the results of Fig.\ref{Fig7}(a), i.e. the maximum charging power $P_{B,\max}$ increases with increasing the $\Omega$. While, the situation is different for $\Delta \neq 0$ and the charging power decreasing with increasing the strength of the classical deriving field, while, the maximum charging power increases with increasing $\Delta$. The reason has been explain in weak coupling section. Fig.\ref{Fig10}(b), represents the maximum stored energy in terms of $\Omega$. As can be seen the maximum stored energy  increases with increasing the $\Omega$ for different values of $\Delta$. It can also be seen that the maximum stored energy increases with increasing $\Delta$. Fig.\ref{Fig10}(c), shows that the ergotropy increases with increasing the $\Omega$ for resonant and non-resonant case. 
\begin{figure}\label{Fig11}
\centering
    \includegraphics[width =1 \linewidth]{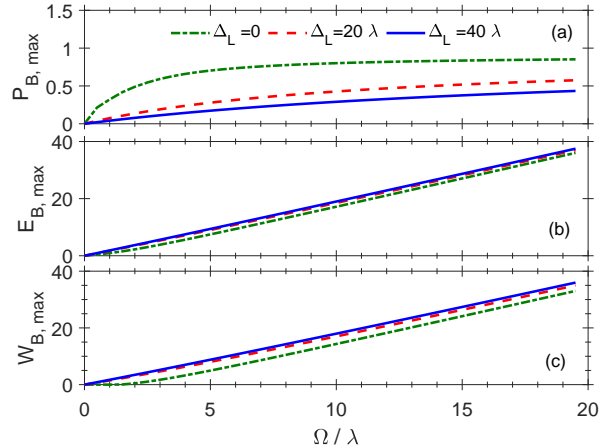}
      \vspace*{-15mm}
    \caption{(a)The maximum charging power $P_{B,\max}$ (b)The maximum internal energy $E_{B,\max}$(c)The maximum Ergotropy $W_{B,\max}$, as a function of the charging time $\tau$ in strong coupling regime with $R=10$ for different values of $\Delta_L$ (the detuning between the classical driving field  and central frequency of
the cavity ), $\Delta=0$ and $r_1=\frac{1}{\sqrt{2}}$. }
    \label{Fig11}
  \end{figure}
In Fig.\ref{Fig11}(a), the maximum charging power has been plotted as a function of   the  classical deriving field strength $\Omega$ for different value of $\Delta_L$ when $\Delta=0$. As can be seen for different value of $\Delta_L$ the maximum charging power increases with increasing the $\Omega$. It can also be seen that increasing $\Delta_L$ decreases the maximum charging power. The maximum stored energy in quantum battery has been plotted in terms of $\Omega$ for different values of $\Delta_L$ . It can be seen that the maximum stored energy increases with increasing $\Omega$ and $\Delta_L$. 
\section{Summary and conclusion}\label{sec4}
In summary, the effects of classical deriving field and its related parameter on performance of quantum battery in leaky cavity have been investigated. We have considered both weak and strong coupling regime. For both resonance mode, that is, when $\Delta=\Delta_L=0$, increasing the strength of the classical field increases the charging power and energy stored in the quantum battery, while it has no effect on the work that can be extracted from the battery, and the amount of ergotropy will always be zero. In complete resonance case the quantum battery can absorb energy from the field, and it will be excited. At the same time, the cavity mode can also be excited to a higher energy level by the driving field. As the driving field becomes stronger, more energy can be transferred to the quantum battery and the cavity, leading to an increase in the charging power and stored energy.

The effect of detuning between classical deriving field and quantum battery $\Delta$ on performance of quantum battery has also been investigated when $\Delta_L=0$. However, when the frequency of the classical field is detuned from the frequency of the quantum battery, the energy exchange is reduced, and the system can operate in a regime where the charging power, stored energy and ergotropy are increased.

We also investigated the effect of detuning between frequency of classical deriving field and central frequency of cavity on performance of quantum battery. The charging power of a quantum battery in a cavity is affected by the detuning between the classical driving field and the central frequency of the cavity due to the phenomenon of resonance.

When the driving field and the cavity are resonant, the energy from the driving field is efficiently transferred to the quantum battery in the cavity, leading to a higher charging power. However, if the driving field and the cavity are detuned, the efficiency of this energy transfer decreases. This is because the detuning causes a phase difference between the driving field and the cavity, resulting in a reduced overlap between the wavefunctions of the driving field and the cavity. As a result, the probability of energy transfer decreases, leading to a lower charging power.



\vfill
\begin{thebibliography}{00}
\bibitem{a1}R. Alicki and M. Fannes, Entanglement boost for extractable
work from ensembles of quantum batteries, Phys. Rev. E 87,
042123 (2013).
\bibitem{a2}T. P. Le, J. Levinsen, K. Modi, M. M. Parish, and F. A. Pollock,
Spin-chain model of a many-body quantum battery, Phys. Rev.
A 97, 022106 (2018).
\bibitem{a3}F. C. Binder, S. Vinjanampathy, K. Modi, and J. Goold, Quantacell:
Powerful charging of quantum batteries, New J. Phys. 17,
075015 (2015).
\bibitem{a4}F. Campaioli, F. A. Pollock, F. C. Binder, L. Céleri, J. Goold, S.
Vinjanampathy, and K. Modi, Enhancing the Charging Power
of Quantum Batteries, Phys. Rev. Lett. 118, 150601 (2017).
\bibitem{a5}G. M. Andolina, D. Farina, A. Mari, V. Pellegrini, V.
Giovannetti, and M. Polini, Charger-mediated energy transfer in exactly solvable models for quantum batteries, Phys. Rev. B
98, 205423 (2018).
\bibitem{a6}I. Henao and R. M. Serra, Role of quantum coherence in the
thermodynamics of energy transfer, Phys. Rev. E 97, 062105
(2018).
\bibitem{a7}J. Q. Quach and W. J. Munro, Using Dark States to Charge and
Stabilize Open Quantum Batteries, Phys. Rev. Appl. 14, 024092
(2020).
\bibitem{a8}F. Pirmoradian and K. Molmer, Aging of a quantum battery,
Phys. Rev. A 100, 043833 (2019).
\bibitem{a9}Y. Huangfu and J. Jing, High-capacity and high-power collective
charging with spin chargers, Phys. Rev. E 104, 024129
(2021).
\bibitem{a10}C.-K. Hu, J. Qiu, P. J. P. Souza, J. Yuan, Y. Zhou, L. Zhang,
J. Chu, X. Pan, L. Hu, J. Li, Y. Xu, Y. Zhong, S. Liu, F. Yan,
D. Tan, R. Bachelard, C. J. Villas-Boas, A. C. Santos, and D.
Yu, Optimal charging of a superconducting quantum battery,
arXiv:2108.04298v1.
\bibitem{a11} W. Lu, J. Chen, L.-M. Kuang, and X. Wang, Optimal state
for Tavis-Cummings quantum battery via Bethe ansatz method,
Phys. Rev. A 104, 043706 (2021).
\bibitem{a12}D. Ferraro, M. Campisi, G. M. Andolina, V. Pellegrini, and
M. Polini, High-Power Collective Charging of a Solid-State
Quantum Battery, Phys. Rev. Lett. 120, 117702 (2018).
\bibitem{a13}A. Crescente, M. Carrega, M. Sassetti, and D. Ferraro, Ultrafast
charging in a two-photon Dicke quantum battery, Phys. Rev. B
102, 245407 (2020).
\bibitem{a14}Y.-Y. Zhang, T.-R. Yang, L. Fu, and X. Wang, Powerful harmonic
charging in a quantum battery, Phys. Rev. E 99, 052106
(2019).
\bibitem{a14p}H.-P. Breuer and F. Petruccione, Theory of Open Quantum
Systems (Oxford University Press, New York, 2002).
\bibitem{a15}F. Barra, Dissipative Charging of a Quantum Battery, Phys. Rev.
Lett. 122, 210601 (2019).
\bibitem{a16}S.-Y. Bai and J.-H. An, Floquet engineering to reactivate a dissipative
quantum battery, Phys. Rev. A 102, 060201(R) (2020).
\bibitem{a17}F. Zhao, F.-Q. Dou, and Q. Zhao, Quantum battery of interacting
spins with environmental noise, Phys. Rev. A 103, 033715
(2021).
\bibitem{a18}S. Gherardini, F. Campaioli, F. Caruso, and F. C. Binder, Stabilizing
open quantum batteries by sequential measurements,
Phys. Rev. Research 2, 013095 (2020).
\bibitem{a19}D. Farina, G. M. Andolina, A. Mari, M. Polini, and V.
Giovannetti, Charger-mediated energy transfer for quantum batteries:
An open-system approach, Phys. Rev. B 99, 035421
(2019).
\bibitem{a20} R. Alicki, A quantum open system model of molecular battery
charged by excitons, J. Chem. Phys. 150, 214110 (2019).
\bibitem{a21}M. T. Mitchison, J. Goold, and J. Prior, Charging a quantum
battery with linear feedback control, Quantum 5, 500
(2020).
\bibitem{a22}S. Seah, M. Perarnau-Llobet, G. Haack, N. Brunner, and S.
Nimmrichter, Quantum Speed-Up in Collisional Battery Charging,
Phys. Rev. Lett. 127, 100601 (2021).
\bibitem{a23}D. Rosa, D. Rossini, G. M. Andolina, M. Polini, and M.
Carrega, Ultra-stable charging of fast-scrambling SYK quantum
batteries, J. High Energy Phys. 11 (2020) 067.
\bibitem{a24}S. Ghosh, T. Chanda, and A. Sen(De), Enhancement in the
performance of a quantum battery by ordered and disordered
interactions, Phys. Rev. A 101, 032115 (2020).
\bibitem{a25}F. H. Kamin, F. T. Tabesh, S. Salimi, F. Kheirandish, and
A. C. Santos, Non-Markovian effects on charging and selfdischarging
process of quantum batteries, New J. Phys. 22,
083007 (2020).
\bibitem{a26}F. T. Tabesh, F. H. Kamin, and S. Salimi, Environment-mediated
charging process of quantum batteries, Phys. Rev. A 102,
052223 (2020).
\bibitem{a27}S.-M. Fei,  N. Jing,  Equivalence of quantum states under local unitary transformations. Physics Letters A 342, 77–81 (2005).
\bibitem{a28}A. Nourmandipour, A. Vafafard, A. Mortezapour,  R. Franzosi, Entanglement protection of classically driven qubits in a lossy cavity. Sci Rep 11, 16259 (2021).
\bibitem{a29}Mortezapour, A. Nourmandipour,  H. Gholipour,  The effect of classical driving field on the spectrum of a qubit and entanglement
swapping inside dissipative cavities. Quantum Inf. Process. 19, 1–16 (2020).
\bibitem{a30}Z. Huang,  H. Situ,  Non-Markovian dynamics of quantum coherence of two-level system driven by classical field. Quantum Inf.
Process. 16, 222, (2017).



\end{thebibliography}
\end{document}